\documentclass[aps,article,tightenlines,showpacs]{revtex4}

\begin{document}


\title{ Circular stationary cyclic symmetric spacetimes; conformal flatness}

\author{Alberto A. Garc\'{\i}a}
 \email{aagarcia@fis.cinvestav.mx}

\author{Cuauhtemoc Campuzano}
 \email{ccvargas@fis.cinvestav.mx}

\affiliation{Departamento~de~F\'{\i}sica,~%
Centro~de~Investigaci\'on~y~de~Estudios~Avanzados~del~IPN\\
Apdo.\ Postal 14-740, 07000 M\'exico DF, MEXICO}

\date{\today}

\begin{abstract}
A subclassification of stationary spacetimes, endowed with one
timelike and one spacelike Killing vectors, i.e., Petrov $G{_2}I$
on $T_2$ spaces, is proposed. Special attention deserves the
Collison's theorem~{\cite{col}} and the branch of metrics
circularly cyclicly (axially) symmetric possessing additionally
the conformal flatness property reported by Garc\'{\i}a and
Campuzano {\cite{ag}}.
\end{abstract}


\maketitle

Although the definition of a cyclicly symmetric spacetime was
introduced by Carter in 1970~{\cite{car}}, the literature on the
topic is rather scarce. It was only until recently that the
concept of cyclic symmetry attracts the attention of researchers
in Relativity: Barnes (2001, 2000)~{\cite{bar}}, Heusler
(1996)~{\cite{heu}}, Mars and Senovilla (1993)~{\cite{sen1}},
Bi$\breve{\rm c}$\'ak and Schmidt (1984)~{\cite{bic}}. Cyclic
symmetry emerges as a slight generalization of the concept of
axial symmetry. Following Carter, a spacetime is called cyclicly
symmetric if and only if the metric is invariant under the action
$\pi$ of the one--parameter cyclic group ${\it SO(\rm 2)}$. A
cyclicly symmetric spacetime is called axisymmetric if the set of
fixed points of $\pi$ is not empty, this set is referred to as the
axis of symmetry. A spacetime is called stationary and
axisymmetric if it is both stationary and axisymmetric, and if the
Killing fields generating the symmetries commute with each other,
see Heusler (1996)~{\cite{heu}}.

On the other hand, the normal form of spacetime metrics admitting
an Abelian group of motions $G_2$ acting on $V_2$ has been
determined by Petrov~\cite{petrov} by means of the metric tensor
of the class $G_2 I$: $g_{ij}=g_{ij}(x^1,x^2)$ endowed with a pair
of commuting Killing vectors ${\it \xi}\equiv{\it
\partial}_3$, and ${\it \eta}\equiv{\it
\partial}_{4}$. The  2-surface of transitivity ( group orbit)
spanned by ${\it \xi}$ and  ${\it \eta}$  is timelike (timelike
group orbits $T_2$) when the square of the simple bivector
$\xi_{[a}\eta{_{b]}}$ is negative (for signature +2), see Kramer et
al.~\cite{kr}.

Stationary cyclicly symmetric spacetimes are those Petrov $G_2 I$
spaces that possess timelike ${\it \xi}\equiv{\it
\partial}_t$ and spacelike ${\it \eta}\equiv{\it
\partial}_{\phi}$ fields of commuting Killing vectors,
such that the trajectories of ${\it
\partial}_{\phi}$ are closed curves, which are referred to as
Petrov $G{_2}I$ on $T_2$ spaces.
The Killing trajectories span timelike
2-surface $T_2$. Therefore, the metric tensor
$g_{ij}=g_{ij}(x^1,x^2)$, $i,j=1,...,4$, possesses, in general,
ten components independent
of the Killing coordinates $t$ and $\phi$, which can be reduced to
six by coordinate transformations.

Circular stationary cyclicly symmetric spacetimes are those ones,
which besides stationarity and cyclic symmetry, exhibit the
simultaneous reflection (inversion in the Chandrasekhar's
terminology~{\cite{chan}}) symmetry
$(t,\phi)\rightarrow(-t,-\phi)$. This invariance yields to the
vanishing of the metric components
$g_{1t}=g_{2t}=g_{1\phi}=g_{2\phi}=0$, and thus the spacetime
splits into two 2-surfaces orthogonal one to the other. The
existence of 2-surfaces orthogonal to the group orbits
(orthogonally transitive group) imposes conditions (circularity
conditions) on the Killing vectors, see ~{\cite{kr}}:
$\xi_{[a;b}\xi_{c}\eta{_{d]}}=0=\eta_{[a;b}\eta_{c}\xi{_{d]}}$.
Most of the exact solutions of physical relevance belong to its
two sub-branches: circular stationary cyclicly symmetric metric
(CSCM) and  circular stationary axisymmetric metric (CSAM)

As subclasses of Petrov $G_2I$ spacetimes one can distinguish:
non--circular stationary cyclicly symmetric metric (N--CSCM) and
non--circular stationary axisymmetric metric (N--CSAM); to our
knowledge, there are no exact solutions belonging to these
subclasses in the literature, thus the search for this kind of
metric is open.

Summarizing, within the class of Petrov $G{_2}I$ on $T{_2}$ spaces
one distinguishes the following  subclasses of metrics:
\noindent CSCM: Circular stationary cyclicly symmetric metric, \\
\noindent CSAM: Circular stationary axisymmetric metric, \\
\noindent N--CSCM: Non--circular stationary cyclicly symmetric metric,\\
\noindent N--CSAM: Non--circular stationary axisymmetric metric.

The class of circular stationary axisymmetric spacetimes, CSAM,
has been studied extensively through the Lewis--Papapetrou metric
and its variations, see Ref.~{\cite{kr}}.

The literature on exact solutions belonging to circular stationary
cyclicly symmetric metrics, CSCM, is rather scarce. Nevertheless,
the general form of conformally flat circular stationary cyclicly
symmetric spacetimes, i.e., a CSCM  supplemented with the property
of conformal flatness, recently has been reported by us in Eq.
(51) of Ref. ~{\cite{ag}}; the explicit form of the line--element
is
\begin{eqnarray*}
ds^2=e^{-2G(x,y)} \bigglb[\frac{dx^2}{
(C_0+C_1x)(x^2+1)}+(C_0+C_1x)dy^2 + (x^2+1){d\phi}^2
       -(dt+x\,d\phi)^2 \biggrb].
\end{eqnarray*}

It is easy to establish that this metric does not possess an axis
of symmetry; accomplishing a $GL(2,R)$ transformation on the
Killingian variables, ${t}$ and ${\phi}$, it occurs that there is
no way to determine an axis of symmetry because of the vanishing
of the determinant of the transformation along any assumed
existing axis.  The cyclic property of the metric above was
pointed out by Barnes and Senovilla~\cite{barsen}, to whom we
acknowledge their contribution to clarify this point (see
\cite{gaca}).

As far as to the general conformally flat {\it circular}
stationary axisymmetric metric is concerned,
Collinson~{\cite{col}} established the following theorem: Every
conformally flat stationary axisymmetric
spacetime is necessarily static.\\
Strictly speaking, the rigorous statement of the Collinson's
theorem should be: Every conformally flat {\it circular}
stationary axisymmetric spacetime is necessarily static. The
adjective {\it circular} should be omitted if the non existence of
non--circular spacetimes of the studied classes is proved; in the
positive case obviously the use of the adjective {\it circular}
should be redundant.

\end{document}